\documentstyle[epsfig,aps,preprint]{revtex}
\tighten
\begin{document}
\draft
\preprint{\vbox{To appear in Physics Letters B \hfill USC-PHYS-NT-01-99}}

\title{Bound state in the vector channel 
of the extended Nambu-Jona-Lasinio model at fixed $f_{\pi}$}
\author{V. Dmitra\v sinovi\' c}
\address{Department of Physics and Astronomy,\\
University of South Carolina, Columbia, SC 29208, USA}
\date{\today}
\maketitle
\begin{abstract}
We show that, as a consequence of fixing $f_{\pi}$ = 93 MeV:
(1) a bound state pole in the
the $J^P = 1^-$ scattering amplitude of the ENJL model exists for
arbitrarily weak (positive) vector coupling $G_2$ so long as the constituent 
quark mass is sufficiently large;
(2) there is a bound state for any quark mass when $G_2 \geq 0.6/(8 f_{\pi}^2 )$;
(3) this bound state becomes massless at $G_2 = 1/(8 f_{\pi}^2 )$ and a 
tachyon for $G_2$ exceeding it. 
We show by way of an 
example that the model has no trouble fitting
the $\rho$ meson mass simultaneously with other observables. 
\end{abstract}

\paragraph*{Introduction.}

Extension of the Nambu and Jona-Lasinio [NJL] model to include vector 
and axial-vector mesons can be traced back to the original paper \cite{njl61}.
The results of this extension concerning vector mesons have undergone considerable 
change
with time: it has long been understood that bound states exist for sufficiently 
strong vector coupling $G_2$, but it is also believed that
such high values of $G_2$ are incompatible with the phenomenology. In the
early 90's Takizawa et al. \cite{tak91} found a new solution to the 
Bethe-Salpeter (BS) equation at lower values of $G_2$, that lay,
however, on the the second lower Riemann sheet of the $J^{\pi} = 1^-$
elastic scattering matrix element.
This pole was interpreted as a ``virtual bound state'', in analogy with the
nonrelativistic situation, which, however, involves only two Riemann sheets 
vs. $\infty$ many present here. 
The extended Nambu-Jona-Lasinio [ENJL] model and its solutions have recently 
come under renewed scrutiny \cite{dkl96,bern97}. In these papers, however,
the solutions to the BS equation at different values of $G_2$ have different values 
of the pion decay constant $f_{\pi}$.

It is the purpose of this Letter to show that the ENJL model results 
regarding the vector- and axial-vector states 
undergo a drastic change when the ``sliding'' pion decay constant $f_{\pi}$ 
is replaced by a fixed one.
In particular, we show that, as a consequence of keeping $f_{\pi}$ fixed,
there is a bound state pole in the
the $J^P = 1^-$ scattering amplitude of the ENJL model for any
vector coupling $G_2 \geq 0$, so long as the 
constituent quark mass $m$ is large enough. We exhibit the dependence of 
the minimal necessary quark mass $m_{\rm min}$ on the vector coupling $G_2$.
When $G_2$ exceeds $0.6/(8 f_{\pi}^2 )$ the vector bound state exists 
for all values of the  quark mass. This bound state becomes massless at 
$G_2 = 1/(8 f_{\pi}^2 )$ and a tachyon for $G_2$ exceeding this value. 
We find no other ``resonances'' in this or the axial-vector channel \cite{bern97}. 
Our results ought to have significant consequences in ENJL-based 
models of electroweak interactions, such as ``technicolour,  top-colour'' etc., 
since there also the scalar bilinear v.e.v. must be kept fixed.

\paragraph*{Conventions and preliminaries}
We shall work in the chiral limit throughout this letter for the sake of
clarity.
Both vector and axial-vector (isovector) currents are conserved in the chiral 
limit and the pion is massless.
The extension to the nonchiral case is straightforward.
The chirally symmetric field theory described by ${\cal L}_{\rm NJL}$ 
\begin{eqnarray} 
{\cal L}_{\rm NJL} = \bar{\psi} \big[{\rm i} {\partial{\mkern -10.mu}{/}} ]\psi
&+& G_{1} \Big[ (\bar{\psi} \psi)^2 + 
  (\bar{\psi} {\rm i} \gamma_5 \mbox{\boldmath$\tau$} \psi)^2 \Big] 
\nonumber \\ 
&-& G_{2} \Big[ (\bar{\psi} \gamma_{\mu} \mbox{\boldmath$\tau$} \psi)^2 + 
  (\bar{\psi} \gamma_{\mu} \gamma_5 \mbox{\boldmath$\tau$} \psi)^2 \Big]\:; 
\label{e:lag2}  
\end{eqnarray}  
in both its original $(G_2=0)$ and extended versions ($G_2\neq 0$)
exhibits spontaneous symmetry breakdown into a  
nontrivial ground state  with constituent quark mass generation 
and a finite quark condensate, when dealt with non-perturbatively.
The non-perturbative dynamics of the model to
leading order in $1/N_c$ are described by two Schwinger-Dyson
[SD] equations: the gap equation and the BS equation. 

The original NJL model
has two free parameters: the positive coupling constant $G_1$ of dimension 
(mass)$^{-2}$ and a regulating cutoff  $\Lambda$ that determines the mass scale.
The gap equation establishes a relation
between  the constituent quark mass $m$ and the two free parameters $G_1$ and 
$\Lambda$. 
This relation is not one-to-one, however: there is a (double) continuum of allowed
$G_1$ and $\Lambda$ values that yield the same nontrivial solution $m$ to the 
gap equation.
Even under the assumption that we know the precise value of $m$, which we don't, 
there is still a great deal of freedom left in the ($G_1, \Lambda$)
parameter space.

Blin, Hiller and Schaden \cite{bhs88} showed how one can eliminate one of 
the two continuum degeneracies by fixing the $G_2 = 0$ value of the pion 
decay constant $f_p = f_{\pi}(G_2 = 0)$ at the observed value 93 MeV. 
Starting from the Goldberger-Treiman
(GT) relation $f_{p} g_{p} = m$ one finds 
\begin{equation}
\left({f_{p} \over m} \right)^{2} = g_{p}^{-2}
= {3 \over{(2 \pi)^{2}}} \sum_{s = 0}^{2} C_{s} \log(M^2_s/m^2)~ ,
\label{e:gt} 
\end{equation}
where the $C_s$ and $M^2_s = m^2 + \alpha_s \Lambda^2$
are the standard parameters of the Pauli-Villars (PV) regularization
scheme \cite{iz80}. 
The result of solving the constraint Eq. (\ref{e:gt}) is a quark mass 
$m$ vs. cutoff $\Lambda$ curve, shown in Fig. 1, 
all points on which satisfy $f_{p}$ = 93 MeV.
One can now select a single point on  this curve by calculating 
an observable that is sensitive to the quark mass $m$, but
not very sensitive to non-chiral corrections, and then fitting the 
aforementioned observable to its experimental value.
One such calculation was carried out in Ref. \cite{lemm95}
with the result $m$ = 225 MeV. Such a procedure completely 
determines the free parameters of the NJL model.
 
Now let $G_2 \neq 0$:
This implies a finite renormalization of the ``bare'' ($G_2 = 0$) pion decay 
constant $f_{p}$ to $f_{\pi}$ and of the constituent quark axial coupling 
$g_A$ \cite{dkl96} according to
\begin{equation}
g_A = \left(1 + 8 G_2 f_{p}^{2}\right)^{-1} = 
\left({f_{\pi} \over f_{p}} \right)^{2} .
\label{e:ga} 
\end{equation}
This leads to the relation 
\begin{equation}
g_A = 1 - 8 G_2 f_{\pi}^{2}
\label{e:ga1} 
\end{equation}
between $g_{A}$ and $G_2$ and $f_{\pi}$, the last of which
is kept constant. An $f_{\pi}$-fixing procedure analogous to the 
one described above now yields a separate $m$ vs. $\Lambda$ curve for 
every value of $g_A$, see Fig. 1.
An important consequence of the relation (\ref{e:ga1}) and of the 
second line of Eq. (\ref{e:ga}) is the inequality $0 \leq g_A \leq 1$.
This imposes a new upper bound on $G_2$:
\begin{equation}
G_2  \leq 1/(8 f_{\pi}^{2})  ~,
\label{e:gaineq} 
\end{equation}
apart from the trivial lower bound $G_2  \geq 0$. $G_2$ values
exceeding the bound imply imaginary values of $g_{p}$ and $f_p$, which in turn 
imply complex cutoff $\Lambda$ and/or mass $m$. Physical interpretation of 
such complex objects is lacking.

We see from Eq. (\ref{e:ga1}) that $G_2$ can be determined from the value 
of the constituent quark axial coupling constant $g_A$, at constant $f_{\pi}$.
One common prescription for estimating $g_A$ is based on the SU(6) 
symmetric nucleon wave function and impulse approximation result for the
nucleon axial coupling
\begin{equation}
g_{A}^{N} = 
{5 \over 3} g_{A} = 1.25 |_{\rm expt.}~,
\label{e:gan} 
\end{equation}
which yields $g_{A} = 0.75$. This procedure is subject to the assumption
that there are no two-quark axial current contributions to the nucleon axial
current matrix element, which assumption is known, however, to be in conflict 
with the chiral symmetry of the model \cite{sato98}. 
Hence we shall use $g_{A} = 0.75$ only as an order of magnitude guide.

Perhaps the most important consequence of the fixed $f_{\pi}$
is the fact that the unrenormalized pseudoscalar
$\pi qq$ coupling $g_{p}$ is a function of $G_2$:  
\begin{equation}
\left({g_{p} \over g_{\pi}} \right)^{2} = 
\left({f_{\pi} \over f_{p}} \right)^{2} =
g_A = 1 - 8 G_2 f_{\pi}^{2}~.
\label{e:gh} 
\end{equation}
This fact is the source of the changes in the vector-channel
spectrum of the ENJL model, to be discussed next.

\paragraph*{Solutions to the BS equation}
The BS equation in the vector/axial-vector channel reads
\begin{eqnarray}
1 + 2 G_2 \Pi_{V,A}(s_{V,A}) = 0
\label{e:bse}
\end{eqnarray}
In order to find the bound state roots $0 \leq s_{V,A} \leq 4 m^2$ 
to these equations 
we require the polarization functions $\Pi_{V,A}$ \cite{dkl96} 
\begin{eqnarray}
\Pi_V(s)&=&
- {2\over 3} g_{p}^{-2} \left[ 2 m^2 [F(s)- 1] + s F(s) \right] 
\nonumber \\
\Pi_{A}(s) &=& \Pi_V(s)+ 4 f^2_{p} F(s)
\label{e:pivanda}
\end{eqnarray}
where
\begin{eqnarray}
F(s) = 1 - \frac{3 g_{p}^2}{2 \pi^2} \{\sqrt{- f}
{\rm Arccot}{\sqrt{-f}} - 1 \}_{PV}
\label{e:emform}
\end{eqnarray}
and $f = 1 - 4m^2/s$. Pauli-Villars (PV) regularization of $F(s)$ has been used.
These $\Pi_{V,A}$ are appropriate when $m$ and $\Lambda$, and hence
also $f_p$ and $g_p$ are fixed. 
That is the parameter-fixing procedure that was used in previous 
solutions of the vector BS Eq. extant in the literature. 
But, then Eq. (\ref{e:ga}) implies that the 
physical pion decay constant $f_{\pi}$ changes with varying $G_2$,  
as noticed in Ref. \cite{bern97}.   

If, on the other hand, we insist on keeping $m$ and $f_{\pi}$ (hence also
$g_{\pi}$) fixed, then  $\Pi_{V,A}(s), F(s)$ in Eqs. (\ref{e:pivanda}),
(\ref{e:emform}) are implicit functions of $G_2$.
This implicit $G_{2}$ dependence can be easily made explicit by using
Eq. (\ref{e:ga}):
\begin{eqnarray}
\Pi_V(s,G_2) &=&
- {2 \over{3 g_{A}}} g_{\pi}^{-2} 
\left[ s + \left(s + 2 m^2\right)[F(s) - 1] \right] 
\nonumber
\\
\Pi_{A}(s,G_2) &=& \Pi_V(s)+ 4 {f^2_{\pi} \over g_{A}} F(s)
\nonumber
\\
F(s,G_2) &=& 1 - \frac{3 g_A g_{\pi}^2}{2 \pi^2} \{\sqrt{- f}
{\rm Arccot}{\sqrt{-f}} - 1 \}_{PV}
~,
\label{e:pi1}
\end{eqnarray}
where we kept $g_A$ as an abbreviation for $1 - 8 G_2 f_{\pi}^{2}$,
for the sake of conciseness.
These $\Pi_{V,A}$ lead to solutions to the BS Eq. (\ref{e:bse}) that are
rather different from what they were with a sliding $f_{\pi}$.

In Fig. 2 we show the numerical solutions to the vector channel BS 
Eq. (\ref{e:bse}) on the physical sheet of the S-matrix for both  
sliding-, and fixed-$f_{\pi}$.
There we also show the Takizawa-Kubodera-Myhrer (TKM)
``virtual bound state'' mass, for both the fixed- and the sliding-$f_{\pi}$.
One sees that: 
(a) the onset of the vector bound state is at substantially lower values of $G_2$
than with a sliding $f_{\pi}$;
(b) the bound state mass drops sharply with increasing $G_2$. In Fig. 2 we have
also shown the analytic approximation to the vector bound state mass
\begin{equation}
m_{V}^{2} = {3 g_{A} g_{\pi}^{2} \over{4 G_{2}}} 
= 6 m^{2} \left({g_{A} \over{1 - g_{A}}}\right) ~.
\label{e:mv} 
\end{equation}
It is manifest from Fig. 2 that Eq. (\ref{e:mv}) is a good approximation 
to the exact result as $g_{A} \to 0$, i.e., as $G_2 \to (8 f_{\pi}^{2})^{-1}$, 
but otherwise consistently overestimates the bound state mass.
According to Eq. (\ref{e:mv}) the bound state ought to dissolve for 
$g_{A} \geq 0.4$, but the exact solution shows that the bound state may exist 
at even higher values of $g_A$, i.e., at lower values of $G_2$, depending on
the value of the constituent quark mass $m$.
In the next section we shall find the range of values of $m = m(\Lambda)$ in which 
a bound state exists for a given $G_2$. 

But first, for the sake of completeness
we discuss the properties of the solutions to the the axial-vector
BS Eq. (\ref{e:bse}). There is only one solution to this equation, at 
$G_2 = 1/(8 f_{\pi}^2)$, on the physical sheet, and none on the ``second'' 
lower sheet. 
The reason for this is that $F(s)$, and hence also $\Pi_{A}(s)$ has an imaginary 
part that does not vanish in the region of interest, i.e., for $s \geq 4 m^2$. 
Solutions to the real part of the axial-vector BS Eq. (\ref{e:bse}) are plotted 
in Fig. 2, together with the analytic approximation
\begin{equation}
m_{A}^{2} = m_{V}^{2} + 6 m^2
= {6 m^{2} \over{1 - g_{A}}} ~.
\label{e:ma} 
\end{equation}
This $m_A^2$ {\it must not} be interpreted as the real part of the resonance 
pole position, because the imaginary part of 
$1 + 2 G_2 \Pi_{A}(s)$ does not vanish anywhere in the mentioned
quadrant of the complex $s$ plane, i.e., there is actually no pole in the
S-matrix element. 
It is curious that although the axial-vector BS equation
does not have the ``virtual bound state'' solution on the second {\it lower} sheet,
there is one such solution on the second-, as well as on each of infinitely 
many {\it upper} Riemann sheets. 
[The branch point $s = 4 m^2$ is a logarithmic one.] 
Physical interpretation of these solutions, if it exists at all, remains 
obscure. We have not found any other solutions either in the vector- or in the
axial-vector channels, in particular we have not found the 
new ``resonance solutions'' of Ref. \cite{bern97}. Our present results do not
change the results and conclusions of Ref. \cite{dkl96} regarding the spectral 
sum rules.

\paragraph*{Minimal quark mass necessary for a vector bound state}
In order to determine the values of $G_2, m$ for which the vector-channel
BS Eq. (\ref{e:bse}) has bound state solutions it is sufficient to consider the
inequality
\begin{eqnarray}
1 + 2 G_2 \Pi_{V}(4 m^{2}) \leq 0~.
\label{e:ineq}
\end{eqnarray}
Using Eq. (\ref{e:pi1}) to find
\begin{eqnarray}
\Pi_V(4 m^2) &=&
- {4 f_{\pi}^{2} \over{3 g_{A}}} 
\left[3 F(4 m^2) - 1 \right] 
\nonumber
\\
&=& 
- {8 f_{\pi}^{2} \over{3 g_{A}}} 
\left[1 + g_A \left(\frac{3 g_{\pi}}{2 \pi}\right)^2 \right] 
~.
\label{e:pi2}
\end{eqnarray}
This and the inequality (\ref{e:ineq}) lead to
\begin{eqnarray}
\left(\frac{3 g_{\pi}}{2 \pi}\right)^2 &\geq&
{5 g_{A} - 2 \over{2 g_{A} (1 - g_A)}} ~,
\label{e:ineq2}
\end{eqnarray}
which is our vector bound state criterium. For $g_A \leq 0.4$ the r.h.s. 
of this inequality is non-positive, i.e., 
the inequality is trivially satisfied and there is a vector bound 
state for all real values of $m$. For $g_A \geq 0.4$ this turns into a
lower bound on the constituent quark mass $m$:
\begin{eqnarray}
m &\geq& m_{\rm min}(g_{A}) = 
f_{\pi} \frac{2 \pi}{3}
\sqrt{{5 g_{A} - 2 \over{2 g_{A} (1 - g_A)}}} ~,
\label{e:ineq3}
\end{eqnarray}
which for $g_A = 0.75$ yields the minimal constituent quark mass of 420 MeV. 
That, in turn gives $m_{\rho} \simeq 840$ MeV, not far from the empirical
770 MeV. Fig. 1 then determines the cutoff $\Lambda =$ 750 MeV. 
This example shows that 
the ENJL model can easily accomodate a bound $q \bar q$ state in the 
$\rho$ channel with realistic constituent quark mass and reasonable $g_A$.

\paragraph*{Conclusion}
We have shown that the ENJL model binds $q \bar q$ states in the vector channel
for arbitrarily small positive values of the coupling constant $G_2$ provided the 
constituent quark mass $m$ is large enough, and for all values of $m$ with 
$G_2 \geq 0.6 (8 f_{\pi}^2)^{-1}$, 
as a consequence of keeping $f_{\pi}$ consistently fixed at 93 MeV. 
This implies that bound vector states can be found at substantially lower 
values of $G_2$ than previously believed and the $\rho$ meson mass easily
reproduced. 
The axial-vector state remains unbound for all allowed values of $G_2$.

Further, we have found that $G_2$ must not exceed $(8 f_{\pi}^2)^{-1}$ if
the vector bound state is not to become a tachyon. This ``critical'' value
of the coupling constant determines a phase transition point, the nature of 
the ``second'' phase being unclear at the moment. The ``first'' phase, with
vector coupling below the critical one,
corresponds to a gauged chiral linear sigma model with 
massive gauge bosons $\rho, A_1$ \cite{gg69}. The exact 
local gauge symmetry is recovered at the critical point at which the vector
gauge boson ($\rho$) becomes massless, whereas $A_1$ keeps a mass of
$\sqrt{6} m$ due to the Higgs mechanism. 

\paragraph*{Acknowledgements}
The author would like to thank K. Kubodera and F. Myhrer for discussions and
comments on the manuscript, and R. H. Lemmer for interesting him in this topic

\begin{figure}
\begin{center}
\epsfig{file=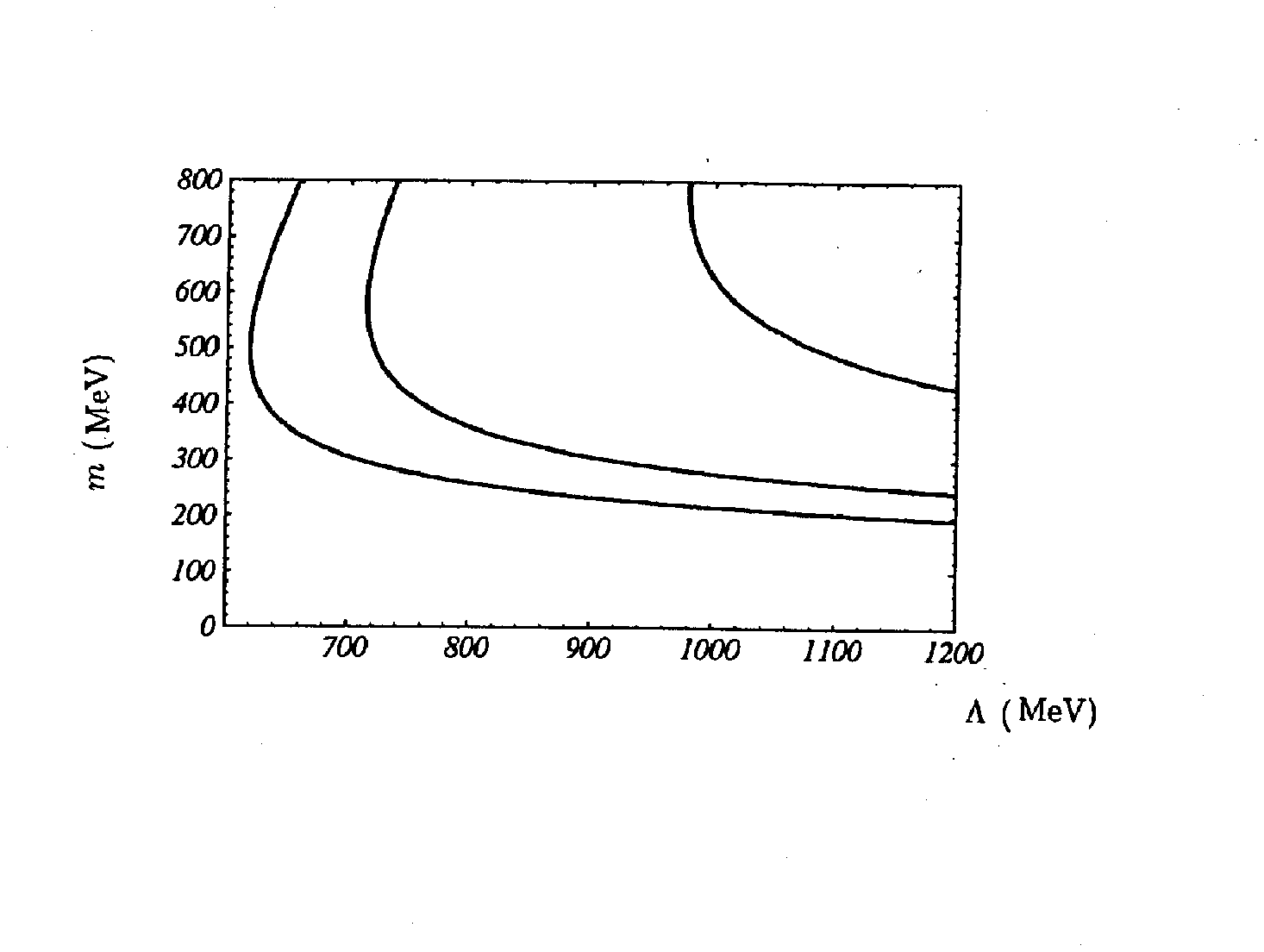,width=16cm} 
\end{center} 
\caption{
The constituent quark mass $m$ as a function of the Pauli-Villars [PV] cutoff 
$\Lambda$ (in units of MeV) 
in the NJL ($g_A = 1$ - the far left h.s. curve), and ENJL models for 
$g_A = 3/4, 2/5$, (the middle and the far right h.s. curves, respectively) 
at fixed $f_\pi = 93$ MeV.}
\label{f:1}
\end{figure}

\newpage
\begin{figure}
\begin{center}
\epsfig{file=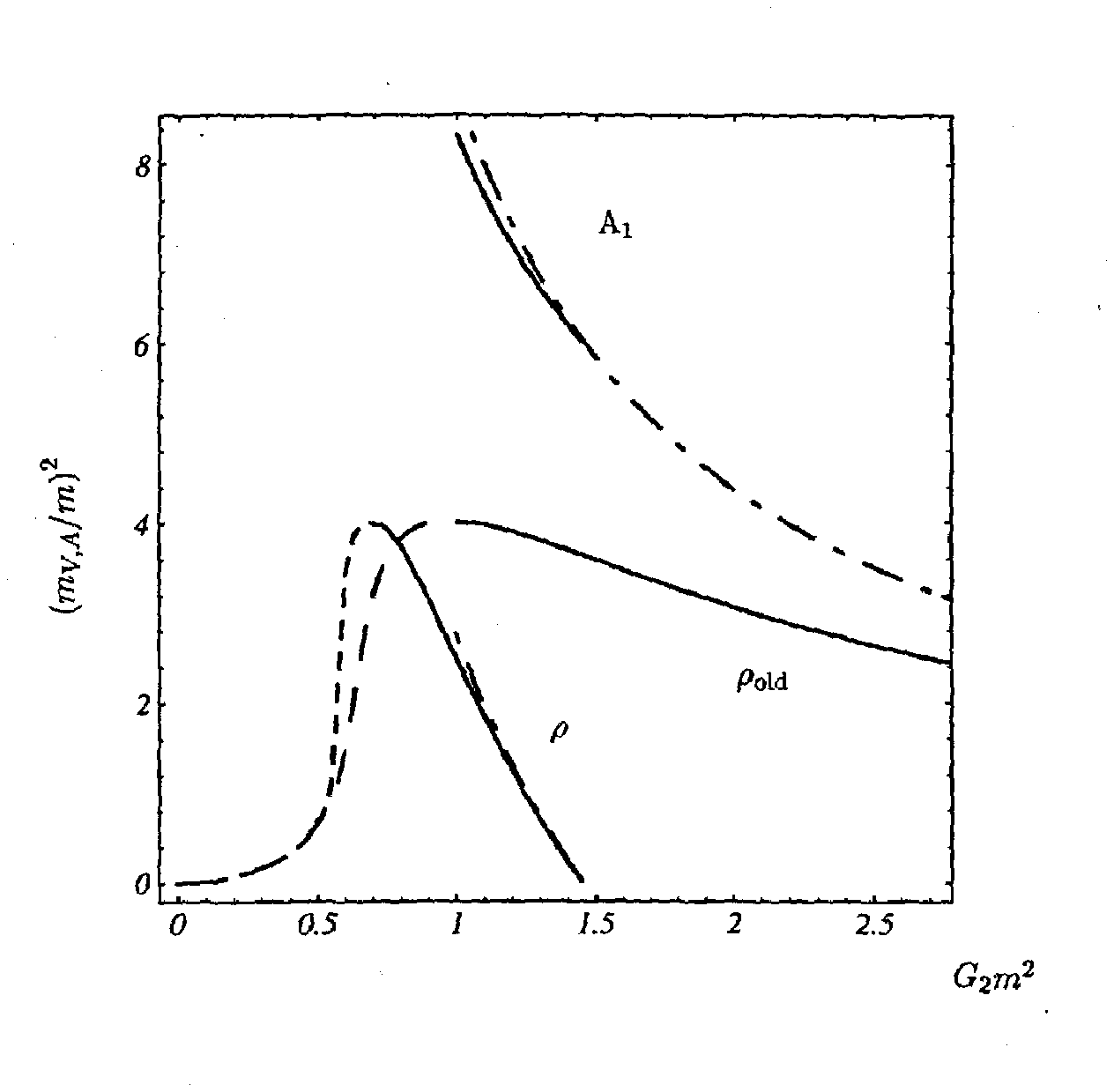,width=16cm} 
\end{center} 
\caption{
Solutions to the BS equation [vector-, or axial-vector 
state mass squared $m_{V, A}^{2}$ rescaled by the constituent quark mass 
squared $m^2$] 
as functions of the rescaled vector interaction coupling constant $G_2 m^2$,
with $m =$ 313 MeV in the ENJL model. [The continuum threshold is at 4.]
(1) vector bound state with sliding $f_{\pi}$ [solid line denoted by
$\rho_{\rm old}$]
continuing into the Takizawa-Kubodera-Myhrer [TKM] ``virtual bound state''  
[long dashes] at lower values of $G_2$; 
(2) vector bound state with fixed $f_\pi$ = 93 MeV [lower solid line denoted by
$\rho$]
continuing into the TKM ``virtual bound state'' with fixed $f_{\pi}$ [short dashes]
at lower values of $G_2$;
(3) root of the real-part of the axial-vector BS Eq. with fixed $f_\pi$ [solid 
line denoted by ${\rm A}_{1}$];
(4) analytic approximations to the vector bound state, and the axial-vector state
at fixed $f_\pi$ [dot-dashes]. }
\label{f:2}
\end{figure}

\end{document}